\documentclass[%
reprint,
superscriptaddress,
bibnotes,
amsmath,amssymb,
aps,
prb,
]{revtex4-2}

\usepackage[main=english,ngerman,french]{babel}
\usepackage[T1]{fontenc}
\usepackage[utf8]{inputenc}
\usepackage{lmodern}
\usepackage{microtype}
\usepackage{graphicx}
\usepackage{hyperref}
\usepackage{amsfonts}
\usepackage{siunitx}
\usepackage{physics}
\usepackage{mathtools}
\usepackage{cleveref}
\usepackage{xfrac}
\usepackage{placeins}
\usepackage{booktabs}
\usepackage{threeparttable}

\makeatletter
\g@addto@macro\bfseries{\boldmath}
\makeatother

\begin{document}

\title{\texorpdfstring{High-Performance Nanophononic Resonators\\in Self-Suspended WSe\textsubscript{2} Domes and Drums}{High-Performance Nanophononic Resonators in Self-Suspended WSe2 Domes and Drums}}

\author{Jens-Christian Drawer}%
\affiliation{Carl von Ossietzky Universität Oldenburg, Fakultät V, Institut für Physik, 26129 Oldenburg, Germany}%

\author{Bo Han}%
\affiliation{Carl von Ossietzky Universität Oldenburg, Fakultät V, Institut für Physik, 26129 Oldenburg, Germany}%

\author{Edson Rafael Cardozo de Oliveira}%
\affiliation{Université Paris-Saclay, C.N.R.S., Centre de Nanosciences et de Nanotechnologies (C2N), 91120 Palaiseau, France}%

\author{Chushuang Xiang}%
\affiliation{Université Paris-Saclay, C.N.R.S., Centre de Nanosciences et de Nanotechnologies (C2N), 91120 Palaiseau, France}%

\author{Vita Solovyeva}%
\affiliation{Carl von Ossietzky Universität Oldenburg, Fakultät V, Institut für Physik, 26129 Oldenburg, Germany}%

\author{Kenji Watanabe}%
\affiliation{Research Center for Electronic and Optical Materials, National Institute for Materials Science, 1-1 Namiki, Tsukuba 305-0044, Japan}%

\author{Takashi Taniguchi}%
\affiliation{Research Center for Materials Nanoarchitectonics, National Institute for Materials Science, 1-1 Namiki, Tsukuba 305-0044, Japan}%

\author{Norberto Daniel Lanzillotti-Kimura}%
\affiliation{Université Paris-Saclay, C.N.R.S., Centre de Nanosciences et de Nanotechnologies (C2N), 91120 Palaiseau, France}%

\author{Christian Schneider}%
\affiliation{Carl von Ossietzky Universität Oldenburg, Fakultät V, Institut für Physik, 26129 Oldenburg, Germany}%

\author{Martin Esmann}%
\thanks{Corresponding author: \href{mailto:m.esmann@uni-oldenburg.de}{m.esmann@uni-oldenburg.de}}
\affiliation{Carl von Ossietzky Universität Oldenburg, Fakultät V, Institut für Physik, 26129 Oldenburg, Germany}%

\date{\today}%

\begin{abstract}
	Van der Waals materials are ideally suited for the implementation of high-frequency nanophononic resonators with atomically flat interfaces. Here, we present two versatile van der Waals-based nanophononic architectures: First, we introduce self-supporting nano-domes of WSe\textsubscript{2} as a scalable platform for the simultaneous generation of hundreds of high-quality nanoacoustic resonators with resonance frequencies in the \SI{100}{\GHz} range. Second, we engineer self-supporting nano-drums that reach record-high working frequencies for 2D-semiconductor transducers beyond \SI{1}{\THz}. Through optical pump--probe spectroscopy experiments and photoelastic linear chain model calculations, we gain a detailed understanding of the intricate interplay between phononic mode hybridization across heterostructures, the differences between modes close to the center and edge of the acoustic Brillouin zone, and the temporal structure of the photoelastic response. Both architectures have potential applications in low-cost nanoacoustic probing and the ultrafast modulation of quantum emitters in two-dimensional semiconductors. While nano-drums surpass the \unit{\THz} frequency barrier, nano-domes appear as an accessible, low-cost alternative for developing scalable nanophononic technologies.
\end{abstract}

\maketitle

\section{Introduction}

Suspended membranes of atomically flat van der Waals (vdW) materials \cite{morell2016} are a versatile optomechanical platform that has been used to study the optical spring effect \cite{xie2021}, radiation-pressure backaction \cite{sanchezarribas2023}, valley-mechanical coupling \cite{li2019}, and nanoscale sensing of thermal transport \cite{morell2019}. Beyond the center of mass motion in membrane geometries, surface acoustic waves (SAWs) have been employed to mechanically modulate the photoluminescence of delocalized excitons in vdW semiconductor layers \cite{nysten2024, peng2022}, which may be extended to single-photon emission centers \cite{buhler2022, choquer2022, weiss2021}. At even higher frequencies, nanophononic engineering of out-of-plane acoustic (ZA) phonons along the $\Gamma$--A direction in vdW slabs and hetero-stacks yields breathing-type resonances in the \SI{10}{\GHz} to \SI{3}{\THz} range \cite{jeong2016, greener2018, soubelet2019, wu2021, zalalutdinov2021, yoon2024, carr2024, aversa2026}, with $Q$-factors up to \num{700} \cite{zalalutdinov2021, yoon2024} if phonon losses are minimized by mechanically suspending the vdW material. Recently, graphene-based heterostructures have enabled the generation, control, and detection of phonons at frequencies reaching \SI{3}{\THz} \cite{yoon2024}, pointing toward useful applications in sensing and phonon filtering. Despite the maturity of vdW stacking techniques, the fabrication of suspended high-performance devices continues to demand deterministic alignment and most importantly transfer onto pre-patterned substrates, limiting throughput and scalability \cite{yoon2024, carr2024, aversa2026}.

Here, we explore a fabrication approach that produces a potentially large number of high-$Q$ nanomechanical resonators in a single process step on a planar substrate. We fabricate nano-domes from the semiconducting transition metal dichalcogenide (TMDC) WSe\textsubscript{2} with lateral dimensions of approximately \SI{1}{\um} and heights up to \SI{50}{\nm}, and demonstrate that these structures remain mechanically intact at cryogenic temperatures. In contrast to previously reported van der Waals nanophononic resonators, the nano-dome architecture offers substantially improved fabrication simplicity and scalability. Using pump--probe spectroscopy \cite{jeong2016, greener2018, soubelet2019, zalalutdinov2021}, we experimentally demonstrate that the domes sustain high-$Q$ breathing resonances beyond \SI{100}{\GHz} at \SI{4}{\K}. By experimental comparison to carefully engineered acoustic drum resonators, we then show that both platforms have comparable performance with $Q$-factors around \num{350}. A linear spring--mass--damper model \cite{yoon2024} reproduces all major spectral and temporal features observed experimentally. By directly comparing dome and drum resonators within a unified experimental and modeling framework, we emphasize the distinct role of vertical heterostructure design in determining the accessible nanophononic bandwidth, extending from \SI{100}{\GHz} to \SI{1.2}{\THz}. For a 2D semiconductor-based transducer, this is the highest confined phonon breathing mode frequency reported so far. Furthermore, nano-domes can be realized through particle irradiation \cite{cianci2023, cianci2022}, rendering the fabrication process fully deterministic and highly scalable. Beyond their role as nanoacoustic resonators, domes and drums both hold promise for low-cost nanoacoustic probing and the ultrafast modulation of quantum emitters in two-dimensional semiconductors \cite{esmann2024, shabani2022, velja2026}.

\section{Methods and results}

\begin{figure*}
	\includegraphics{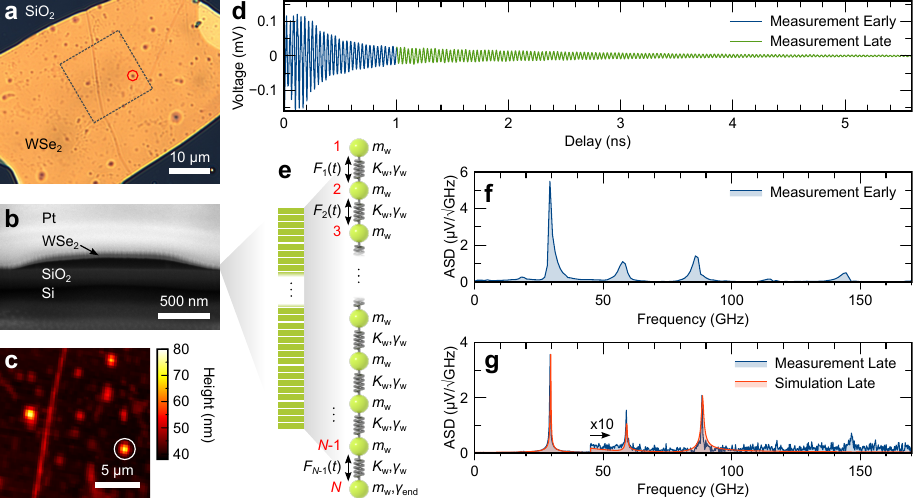}%
	\caption{\label{fig:fig1}%
		(a)~Microscope image of the WSe\textsubscript{2} with nano-domes. (b)~SEM image of a single nano-dome dissected by focused ion beam lithography after deposition of a thin platinum layer. (c)~AFM scan of the sample area highlighted as a dashed box in (a). A height of \SI{0}{\nm} corresponds to the SiO\textsubscript{2} substrate surface. A white circle marks the dome investigated by pump--probe spectroscopy. (d)~Measured differential reflectivity time trace on the dome after background subtraction. (e)~Linear mass--spring--damper chain model used to calculate the theoretical spectrum in panel~(g). (f)~Pump--probe spectrum obtained by Fourier transform of the time trace's time derivative for a temporal window of \qtyrange{10}{1000}{\ps} (``early'') and (g) \qtyrange{1}{5.6}{\ns} (``late'', blue). The spectrum in panel~(g) was magnified $\times10$ above \SI{45}{\GHz} for better visibility. Panel~(g) also shows the simulation ``late'' (red) obtained from the model in panel~(e), which is in excellent agreement with the measured curve.}%
\end{figure*}

The first sample consists of a single WSe\textsubscript{2} flake of \SI{43.5}{\nm} thickness that we transferred onto a planar SiO\textsubscript{2}/Si substrate by dry stamping \cite{castellanos-gomez2014}. By choosing a high speed for the dry transfer process around \SI{50}{\um\per\s} defined as the lateral propagation velocity of the contact point between the 2D material and the substrate, a large number of micrometer-sized air bubbles can be trapped between them. This leads to the formation of nano-domes with lateral diameters in the range of \qtyrange{0.5}{2.0}{\um} with a corresponding height of up to \SI{50}{\nm} \cite{khestanova2016, sanchez2021, du2025}. \Cref{fig:fig1}(a) shows an optical microscope image of the sample. One of the domes that we later investigated by pump--probe spectroscopy is marked with a red circle. \Cref{fig:fig1}(b) shows a scanning electron microscope (SEM) image of a single dome that we dissected by focused ion beam milling. The elevated WSe\textsubscript{2} membrane and the micron-sized void below are clearly identified. The dome was coated with a thin layer of Platinum prior to dissection to allow for a sharper ion beam cut. Nano-domes formed all over the TMD flake with typical distances between them of few microns, i.e., we produced around \num{100} nanoresonators in a single step transfer process. To precisely determine the lateral extent and height of the domes, we performed an atomic force microscopy (AFM) scan of the relevant sample region marked with a dashed box in panel~(a). The scan displayed in panel~(c) shows a height distribution from \qtyrange{2}{40}{\nm} above the WSe\textsubscript{2} slab height. 

To probe the nanophononic response of the nano-dome encircled in panels~(a, c), we performed local pump--probe spectroscopy at a temperature of \SI{3.8}{\K}. The nano-domes stay suspended at these temperatures and show no signs of degradation upon thermal cycling. For a detailed description of the optical setup, refer to section~I of the Supporting Information. We use a Ti:Sapphire laser emitting pulses with \SI{140}{\fs} nominal duration at a repetition rate of \SI{80}{\MHz} and \SI{732}{\nm} central wavelength tuned to the low-energy flank of the free intralayer exciton absorption dip of WSe\textsubscript{2}. By passing the laser through a polarizing beam splitter (PBS), we prepare two orthogonally polarized pulse trains and send one through a motorized mechanical delay line. The relative power of the two pulses in each pair is controlled with a $\sfrac{\lambda}{2}$-waveplate in front of the PBS. The two beams are recombined in a second PBS and focused tightly onto the dome to an estimated focal diameter of \SI{1}{\um}. In this scheme, the first pulse acts as the pump and launches a coherent pulse of longitudinal acoustic phonons into the WSe\textsubscript{2} membrane of the dome. The second pulse acts as a probe and detects transient changes in optical reflectivity caused by the instantaneous mechanical strain distribution of the phonon pulse. The reflected light is passed through a set of waveplates and a polarizer, allowing for suppression of the pump pulse, before the reflected probe impinges onto a fast photo diode. We modulate the pump at a frequency of \SI{1}{\MHz} by an acousto-optical modulator and detect pump-induced changes in reflectivity via lock-in demodulation of the photo diode signal.

\Cref{fig:fig1}(d) displays the transient reflectivity for pump--probe delays up to \SI{5}{\ns} where an incoherent background from electronic contributions to the signal was already subtracted to highlight the coherent contributions \cite{ruello2015}. We analyze the spectral composition of this trace by calculating the Fourier transform of its first time derivative, which is shown in panels~(f, g). We find major contributions to the spectrum at \SI{30}{\GHz}, \SI{60}{\GHz}, and \SI{90}{\GHz}, which we assign to the first three breathing type longitudinal acoustic phonon modes of the WSe\textsubscript{2} dome. These frequencies are in good agreement with the measured flake thickness of \SI{43.5}{\nm} and a literature value for the out-of-plane speed of sound in WSe\textsubscript{2}, i.e., the ZA phonon dispersion along the $\Gamma$--A direction, of \SI{2.5}{\km\per\s} \cite{vialla2020}. We note that the measured quality factors differ substantially when either including only the trace up to \SI{1}{\ns} (f) or the remaining trace between \qtyrange{1}{5.6}{\ns} (g) in the Fourier transform. In both cases, the resulting linewidths are substantially larger than the frequency resolution limit imposed by the length of the Fourier transform window. Therefore, this difference in quality factor is likely caused by simultaneous probing of the dome and parts of the surrounding flake in contact with the substrate, since our optical focus is slightly larger than the dome. Since the phonon lifetime outside the dome is significantly shorter than for the suspended membrane, the measured $Q$-factors of \num{227\pm11} (\SI{30}{\GHz} mode) and \num{330\pm50} (\SI{90}{\GHz} mode) at larger delays (panel~(g)) contain only contributions from the suspended part. $Q$-factors were extracted from this amplitude spectrum by fitting the square root of a Lorentzian to each peak as shown in fig.~S2 of the Supporting Information. To underpin the assignment of these two contributions, we show a reference pump--probe spectrum measured next to the dome as fig.~S3 in the Supporting Information. The $Q$-factor of the \SI{30}{\GHz} mode is found to be \num{10.4\pm1.3}, similarly low as in \cref{fig:fig1}(f). We quantitatively model the pump--probe spectrum at large delay by describing the acoustic properties of the sample as a chain of masses connected by springs (panel~(e)), where each mass models one atomically thin monolayer of the (vdW) crystal (see section~II of the Supporting Information for technical details and parameters). We use established literature values for mass densities $m_\textrm{w}$, layer thicknesses and spring constants $K_\textrm{w}$. The coherent excitation is implemented as an ultrashort Gaussian force profile $F_i(t)$ driving the masses apart at time $t=0$. For each pair of masses along the chain the force is set proportional to the optical intensity of the incident laser pulse \cite{thomsen1986, lanzillotti-kimura2007}. We let the system evolve in time and model the transient reflectivity at times $t>0$ as the sum over the strain between neighboring layers weighted with the optical mode profile of the laser obtained from transfer matrix simulations. Keeping the damping terms in the chain $\gamma_\textrm{w}$ as the only essential fit parameter of the problem, we find excellent agreement between experiment and theory for the suspended WSe\textsubscript{2} dome. We reach this level of agreement despite the fact that our measurement averages over the full dome including regions close to the border that may be pre-strained by several percent \cite{cianci2022, shabani2022, velja2026}.

\begin{figure*}
	\includegraphics{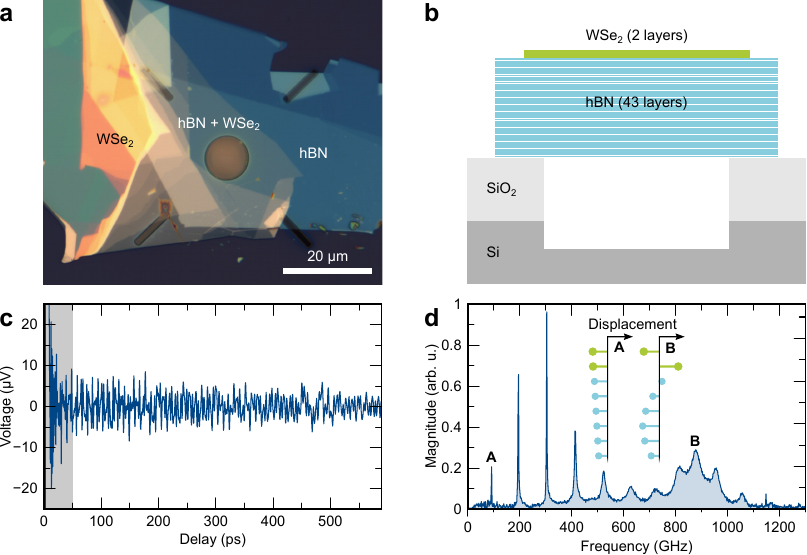}%
	\caption{\label{fig:fig2}%
		(a)~Microscope image of the investigated sample. The WSe\textsubscript{2} bilayer--hBN stack covers a hole with a diameter of \SI{10}{\um}. (b)~Sketch of the layer structure. (c)~Measured differential reflectivity time trace. The first \SI{50}{\ps} interval is analyzed in more detail in \cref{fig:fig3}. (d)~Measured spectrum corresponding to (c). The insets show the mode shapes obtained as eigenmodes of the linear mass--spring--damper chain model of (b) for the first (A) and ninth (B) modes. The first two green-highlighted layers are the WSe\textsubscript{2} bilayer, which shows in-phase (A) and out-of-phase (B) movement.}%
\end{figure*}

As a second architecture, we study the performance of the established drum design that is usually applied when suspending atomically thin semiconductor flakes \cite{morell2016, xie2021, sanchezarribas2023, li2019, morell2019, zalalutdinov2021, yoon2024}. With this approach we reach the highest transduction frequencies for any 2D semiconductor reported so far, reaching a confined breathing mode resonance at \SI{1.2}{\THz}. In \cref{fig:fig2}(a) we show a WSe\textsubscript{2} bilayer on a thick slab of hexagonal Boron nitride (hBN) support, both suspended over a \SI{10}{\um} diameter hole of \SI{1}{\um} depth, which we prepared by focused ion beam milling of the Si/SiO\textsubscript{2} substrate before deterministically dry-stamping the two-dimensional materials. Panel~(b) displays a schematic of the vertical cross-section through the sample. We chose this heterostructure as the simplest non-trivial case of a sample where phononic mode hybridization strongly modifies the acoustic response \cite{zalalutdinov2021}. Next, we performed pump--probe spectroscopy in the exact same manner as on the domes and obtain the pump--probe traces shown in panel~(c). Again, the incoherent background was subtracted. The pump--probe spectrum obtained via a Fourier transform is displayed in panel~(d). Since the drum is substantially larger than our optical focus, we observe no mixing of signals from areas with different $Q$-factors. The spectrum features two main groups of peaks, one with high $Q$-factors and an envelope centered at \SI{300}{\GHz}, the other with lower $Q$-factors and its envelope around \SI{850}{\GHz} (see fig.~S4 of the Supporting Information for a detailed $Q$-factor analysis). In the first group of peaks, we find a maximum $Q$-factor of \num{375}, i.e., in a range very similar to the nano-dome. We assign the two groups of peaks to two different kinds of hybrid nanophononic modes: Since hBN is optically transparent, both phonon generation and detection are almost completely localized to the WSe\textsubscript{2} bilayer. The ZA phonon dispersion of WSe\textsubscript{2} in out-of-plane direction, i.e., along the $\Gamma$--A direction, has its upper band edge at \SI{840}{\GHz} \cite{jeong2016}, which corresponds to the out-of-phase motion in a bilayer, whereas the band edge for ZA phonons in hBN at the A point of the Brillouin zone is much higher around \SI{2.4}{\THz} \cite{serrano2007}. Consequently, at low frequencies the hybrid modes of the drum structure feature near in-phase motion of neighboring (vdW) layers. For frequencies around \SI{800}{\GHz} the hybrid modes are the coupled motion of Bloch (plane) waves in hBN and the out-of-phase motion of the WSe\textsubscript{2} bilayer. For the modes labeled A and B we show the corresponding mode profiles as insets to panel~(d). This assessment is corroborated by the observed $Q$-factors and mode spacings. Since phonon damping is roughly an order of magnitude bigger for relative motion of WSe\textsubscript{2} layers than for hBN \cite{yoon2024}, the higher order breathing modes in the second group of peaks feature a low $Q$-factor that is limited by the out-of-phase motion in WSe\textsubscript{2} and varies only little with frequency (see fig.~S4 of the Supporting Information). The ZA phonon cut-off in WSe\textsubscript{2} at \SI{840}{\GHz} furthermore causes a drop in the free spectral range of the observed breathing modes as the group velocity decreases towards the edge of the acoustic Brillouin zone. For frequencies above the band edge around \SI{1}{\THz} the free spectral range is again dictated by the speed of sound in hBN with a constant reflection phase adding to the round trip at the hBN-WSe\textsubscript{2} interface (see fig.~S5 of the Supporting Information).

\begin{figure*}
	\includegraphics{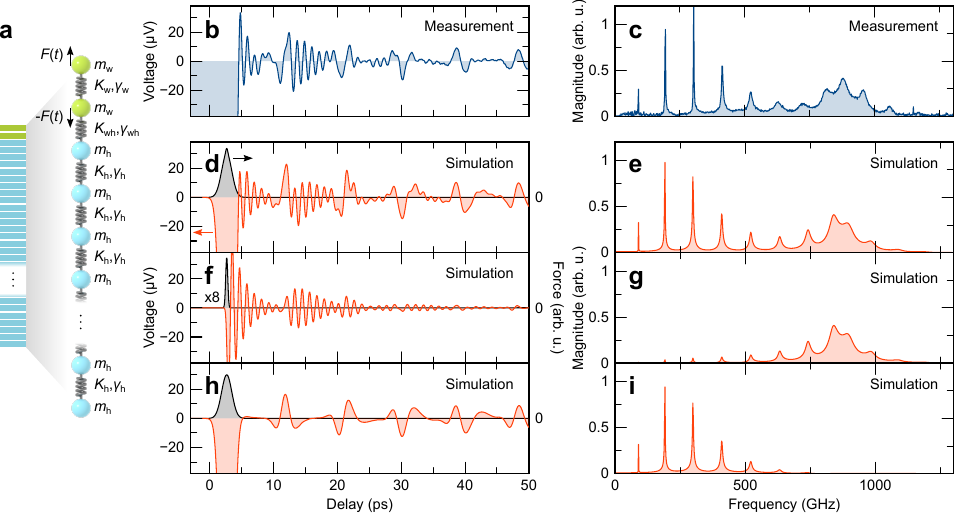}%
	\caption{\label{fig:fig3}%
		(a)~Schematic structure of the linear mass--spring--damper chain. (b)~Detailed plot of the measured differential reflectivity time trace shown in \cref{fig:fig2}(c) for the first \SI{50}{\ps} interval and (c) the corresponding spectrum for the full interval. In (b), the slow background of the measured signal was removed by fitting and subtracting a polynomial of degree \num{20} in the range \qtyrange{5}{55}{\ps}. (d)~Simulated differential reflectivity time trace (orange) for the combination of the slow and fast force components (black) and (e) its corresponding spectrum. (f)~Simulated differential reflectivity time trace (orange) for the fast force component (black) and (g) its corresponding spectrum. (h)~Simulated differential reflectivity time trace (orange) for the slow force component (black) and (i) its corresponding spectrum.}%
\end{figure*}

We quantitatively model the pump--probe time trace and spectrum in \cref{fig:fig2}(c, d) using the same model of masses connected with springs as before. The results are displayed in \cref{fig:fig3} with parameters summarized in section~II of the Supporting Information. We use literature values for the mass densities $m_\textrm{w}$ (WSe\textsubscript{2}) and $m_\textrm{h}$ (hBN), for the force constant between layers for the two materials $K_\textrm{w}$, $K_\textrm{h}$ and the hetero-interface $K_\textrm{hw}$ \cite{yoon2024}. Leaving the damping term $\gamma_\textrm{w}$ identical to before, the only remaining free parameters are the damping in hBN $\gamma_\textrm{h}$ and in the hetero-interface $\gamma_\textrm{hw}$. \Cref{fig:fig3}(a) shows the setup of the model. The initial force pulse exclusively acts on the WSe\textsubscript{2} bilayer, generating phonons via resonant pumping of the TMDC exciton. Since we detect at the same wavelength, we model the detected time trace as the strain in the WSe\textsubscript{2} bilayer, followed by a time derivative and a Fourier transform to obtain corresponding spectra. Panels~(b, c) show a zoom into the experimentally measured time trace taken from \cref{fig:fig2}(c, gray shaded area) and the measured pump--probe spectrum taken from \cref{fig:fig2}(d). The simulated time trace and its Fourier transform are displayed in \cref{fig:fig3}(d, e). We find excellent agreement with the measurements reproducing even fine details of the time trace as well as all main features of the spectrum. Our modeling results reveal that at least two different time scales must be involved in the phonon generation mechanism to account for our experimental findings, in contrast to previous works \cite{yoon2024, carr2024, aversa2026}. These two scales result in the bimodal envelope of the pump--probe spectrum shown in panel~(e). In \cref{fig:fig3}(f), only the shorter time scale is included in the simulation, a Gaussian force profile with $\sigma_\textrm{s}=\SI{160}{\fs}$, which matches the duration of the pump laser pulse. The resulting fast oscillations in the time trace correspond to the high-frequency part of the bimodal spectrum in panel~(g). Including only the other time scale in the excitation force profile (Gaussian with $\sigma_\textrm{l}=\SI{800}{\fs}$) results in the time trace and spectrum shown in panels~(h--i), i.e., the low-frequency part of the overall spectrum. This slow time scale most likely represents the characteristic duration of thermal expansion in the bilayer \cite{yoon2024}, since the pump pulse resonantly excites free intralayer excitons in the WSe\textsubscript{2}.

\begin{table*}
	\caption{\label{tab:comparison}%
		Comparison of the properties of our nano-dome platform to the nano-drum design and other suspension designs from literature. Summarized are lateral size $d_{xy}$, thickness $d_{z}$, maximum $Q$-factor $Q_\textrm{max}$ with the corresponding frequency given in brackets (in \unit{\GHz}), maximum breathing mode resonance frequency $f_\textrm{max}$, and maximum $Qf$ product $(Qf)_\textrm{max}$.}%
	\begin{threeparttable}
		\begin{tabular}{lcccS[table-format=1.1]S[table-format=1.2e2]c}
			\toprule
			\textbf{System} & {\textbf{$d_{xy}$\,(µm)}} & {\textbf{$d_{z}$\,(nm)}} & {\textbf{$Q_\textrm{max}$}} & {\textbf{$f_\textrm{max}$\,(THz)}} & {\textbf{$(Qf)_\textrm{max}$\,(Hz)}} & {\textbf{Scalable}} \\ \midrule
			Dome, WSe\textsubscript{2}, this work              & 0.5--2.0 &   43.5   &      330 (90)      & 0.15 & 2.97e13 & Yes \\
			Drum, WSe\textsubscript{2}/hBN, this work          &    10    &   16.8   &     375 (300)      & 1.1  & 1.12e14 & No  \\
			Drum, MoSe\textsubscript{2} \cite{soubelet2019}    &    6     & 1.3--337 &      190 (43)      & 0.97 & 2.31e13 & No  \\
			Drum, MoS\textsubscript{2} \cite{zalalutdinov2021} &    10    & 2.5--62  &     730 (146)      & 0.6  & 1.1e14  & No  \\
			Opal, WSe\textsubscript{2} \cite{carr2024}         &   N/A    &  8--130  &      439 (34)      & 0.17 & 2.70e13 & Yes \\
			Drum, WSe\textsubscript{2}/hBN/FLG \cite{yoon2024} &  20--25  &    22    &     590 (1800)     & 2.1  & 1e15    & No  \\
			Drum, MoS\textsubscript{2} \cite{aversa2026}       &  5--10   & 10--500  & 1260 (17)\tnote{a} & 0.32 & 2.18e13 & No  \\
			\bottomrule
		\end{tabular}
		\begin{tablenotes}
			\item[a] Measured with time between successive excitation pulses shorter than the reported lifetime \cite{aversa2026}.
		\end{tablenotes}
	\end{threeparttable}
\end{table*}

\Cref{tab:comparison} shows a detailed comparison of the nano-domes introduced in our work to the drum geometry. Our results show that both perform on par with state of the art suspended nanoresonators based on vdW materials hosting breathing type phonon modes \cite{soubelet2019, zalalutdinov2021, yoon2024}. For the nano-domes, we find a maximum $Q$-factor of \num{330} for the mode at \SI{90}{\GHz} that results in a $Qf$ product of \SI{2.97e13}{\Hz}. In the drum resonator, we find a maximum $Q$-factor of \num{375} for the mode at \SI{300}{\GHz} that results in a $Qf$ product of \SI{1.12e14}{\Hz}, i.e., in the range necessary for room-temperature quantum optomechanics \cite{aspelmeyer2014} for both structures. The observed maximum $Q$-factors are close to the resolution limit of our experimental setup, dictated by the length of the pump probe time trace. We note that the reported $Q$-factors may therefore even slightly underestimate the actual performance of the devices under study, in particular for the very low frequency resonances. 

The main difference in the two types of devices studied here is the width of their spectral response with an upper frequency cut-off in intensity around \SI{50}{\GHz} for the nano-dome and around \SI{1.1}{\THz} for the drum. This difference is a direct consequence of the vertical structure in each resonator. In the simplest approximation, the response of a vdW phonon transducer for ZA phonons is centered at the frequency corresponding to its half-wavelength resonance \cite{zalalutdinov2021}, i.e., $f_\textrm{c}=\frac{v_\textrm{vdW}}{2d_\textrm{vdW}}$ with $v_\textrm{vdW}$ the out-of-plane speed of sound and $d_\textrm{vdW}$ its thickness. Assuming a bandwidth similar to the center frequency \cite{zalalutdinov2021}, the cut-off in the case of the nano-dome is predicted at $f_\textrm{cut}\approx0.75\cdot\frac{\SI{2.5}{\km\per\s}}{\SI{43.5}{\nm}}=\SI{43.1}{\GHz}$, matching the experimental result. For the drum, the relevant transducer thickness is the bilayer WSe\textsubscript{2} with a half-wavelength resonance at \SI{840}{\GHz} and a corresponding prediction for the cut-off at $f_\textrm{cut}\approx\SI{1.26}{\THz}$, again in reasonable agreement with the experimental findings. The presence of the bimodal envelope in the frequency response of the drum resonator thus clearly hints at the presence of phonon generation mechanisms beyond a quasi-instantaneous force pulse uniformly acting across the whole vdW transducer. Importantly, the key differences between the nano-dome and drum resonators outlined above are not fundamentally rooted in their transversal structure but solely hinge on the chosen vertical stacking geometry of the vdW materials. Using established dry stamping approaches for the generation of hBN-encapsulated samples and other vdW heterostructures, the nano-dome geometry introduced here can be engineered for higher phonon frequencies or vertically separated phonon generation and detection in vdW materials with exciton resonances at different energies.

\section{Conclusion}

We have presented self-suspended WSe\textsubscript{2} nano-domes as high quality, low-cost na\-no\-pho\-no\-nic resonators that are easy to fabricate in large numbers without any need for sample pre-patterning. Featuring $Q$-factors up to \num{350} at lateral sizes in the \SI{1}{\um} range, these resonators perform on par with state of the art suspended drum resonators, while their footprint pushes the minimum possible area of a resonator that can be probed from the optical far field. Nano-drums, on the contrary, require careful fabrication protocols, but excel in terms of reachable maximum frequency, surpassing the \unit{\THz} barrier. By comparing optical pump--probe spectra of nano-domes and drum resonators, we conclusively show that both exhibit frequency combs of localized breathing-type ZA phonon modes, whose frequencies are controlled by the overall resonator thickness and material composition. The experimental results fully agree with photoelastic model calculations based on a linear mass and spring chain model with a minimum number of free parameters. Our simulations also reveal an interesting bimodal envelope of pump--probe spectra in the case of a TMD bilayer interfaced with hBN, which hints at the possibility of two different time scales being involved in the phonon generation mechanism. In conclusion, our findings underpin vdW materials as an ideally suited material platform for the versatile implementation of nanophononic resonators with resonance frequencies up to the \unit{\THz} range and with potential applications in nanoacoustic probing as well as the ultrafast modulation of quantum emitters in two-dimensional semiconductors. Present developments in nanofabrication techniques point at the possibility of new nanophononic architectures \cite{xu2022}, for example, optimized substrates, such as artificial opal \cite{carr2024}, self-assembly \cite{kanjanaboos2013, abdala2020}, 3D lithography \cite{stassi2021, meier2025}, and the use of commercially available substrates for other applications such as TEM grids and SiN membranes. A particularly interesting perspective is the deterministic generation of nano-domes via hydrogenization of TMDs that may simultaneously result in single photon emitters strain-coupled to the acoustic modes of deterministically placed domes \cite{cianci2023, cianci2022}.

\begin{acknowledgments}
	This project was funded by the \begin{otherlanguage}{ngerman}Deutsche Forschungsgemeinschaft\end{otherlanguage} {} (DFG, German Research Foundation), grant number INST 184/234-1 FUGG. This project was funded within the QuantERA II programme that has received funding from the European Union’s Horizon 2020 research and innovation programme under Grant Agreement No.\ 101017733, and with funding organization the Germany Federal ministry of research, technology and aeronautics within the project EQUAISE. The authors acknowledge support by the European Commission (ERC Dual-Twist, Grant number 101170213 and ERC T-Recs, Grant number 101045089) and through the \begin{otherlanguage}{ngerman}Niedersächsisches Ministerium für Wissenschaft und Kultur\end{otherlanguage} {} (``DyNano'' and Wissenschaftsraum ElLiKo within the programme ``zukunft.niedersachsen''). K.W.\ and T.T.\ acknowledge support from the CREST (JPMJCR24A5), JST and World Premier International Research Center Initiative (WPI), MEXT, Japan. The authors thank Henri Melchert, Keno Bischoff, and Paul Haferkorn for supporting the sample preparation. The authors thank Rebeca Ribeiro-Palau for fruitful discussions about sample design.
\end{acknowledgments}

\bibliography{local}

\end{document}


\title{\texorpdfstring{Supporting Information\\High-Performance Nanophononic Resonators\\in Self-Suspended WSe\textsubscript{2} Domes and Drums}{Supporting Information -- High-Performance Nanophononic Resonators in Self-Suspended WSe2 Domes and Drums}}

\author{Jens-Christian Drawer}%
\affiliation{Carl von Ossietzky Universität Oldenburg, Fakultät V, Institut für Physik, 26129 Oldenburg, Germany}%

\author{Bo Han}%
\affiliation{Carl von Ossietzky Universität Oldenburg, Fakultät V, Institut für Physik, 26129 Oldenburg, Germany}%

\author{Edson Rafael Cardozo de Oliveira}%
\affiliation{Université Paris-Saclay, C.N.R.S., Centre de Nanosciences et de Nanotechnologies (C2N), 91120 Palaiseau, France}%

\author{Chushuang Xiang}%
\affiliation{Université Paris-Saclay, C.N.R.S., Centre de Nanosciences et de Nanotechnologies (C2N), 91120 Palaiseau, France}%

\author{Vita Solovyeva}%
\affiliation{Carl von Ossietzky Universität Oldenburg, Fakultät V, Institut für Physik, 26129 Oldenburg, Germany}%

\author{Kenji Watanabe}%
\affiliation{Research Center for Electronic and Optical Materials, National Institute for Materials Science, 1-1 Namiki, Tsukuba 305-0044, Japan}%

\author{Takashi Taniguchi}%
\affiliation{Research Center for Materials Nanoarchitectonics, National Institute for Materials Science, 1-1 Namiki, Tsukuba 305-0044, Japan}%

\author{Norberto Daniel Lanzillotti-Kimura}%
\affiliation{Université Paris-Saclay, C.N.R.S., Centre de Nanosciences et de Nanotechnologies (C2N), 91120 Palaiseau, France}%

\author{Christian Schneider}%
\affiliation{Carl von Ossietzky Universität Oldenburg, Fakultät V, Institut für Physik, 26129 Oldenburg, Germany}%

\author{Martin Esmann}%
\thanks{Corresponding author: \href{mailto:m.esmann@uni-oldenburg.de}{m.esmann@uni-oldenburg.de}}
\affiliation{Carl von Ossietzky Universität Oldenburg, Fakultät V, Institut für Physik, 26129 Oldenburg, Germany}%

\date{\today}%

\maketitle

\renewcommand{\thefigure}{S\arabic{figure}}
\renewcommand{\theequation}{S\arabic{equation}}

\section{\label{sec:SIpumpprobe}Experimental methods -- Pump--probe spectroscopy}
\FloatBarrier

\begin{figure*}[ht]
	\includegraphics{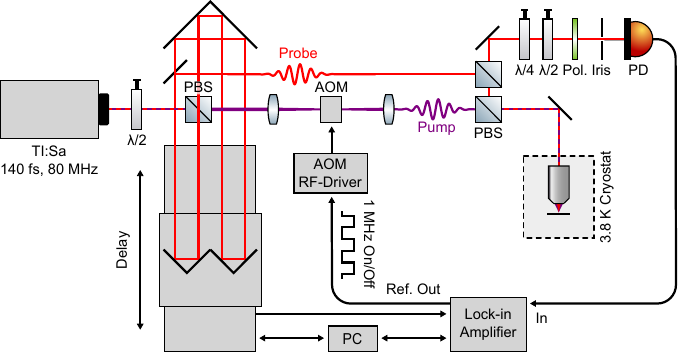}%
	\caption{\label{fig:figS1}%
		Schematic of the pump--probe setup.}%
\end{figure*}

A sketch of the experimental pump--probe setup is shown in \cref{fig:figS1}. The sample is held at \SI{3.8}{\K} in a closed-cycle cryostat (AttoCube AttoDry800) on an XYZ translation stage. Optical imaging as well as cross-polarized optical pump--probe spectroscopy are all performed through the same cold microscope objective (AttoCube LT-APO/VISIR/0.82 with $\mathrm{NA}=\num{0.82}$). We derive optical pulses with \SI{140}{\fs} nominal duration at a repetition rate of \SI{80}{\MHz} and \SI{732}{\nm} central wavelength from a Ti:Sapphire laser (Coherent Chameleon Ultra II). By passing the laser through a polarizing beam splitter (PBS), we prepare two orthogonally polarized pulse trains. The relative power of the two pulse trains is controlled with a $\sfrac{\lambda}{2}$-waveplate in front of the PBS. We send the train of probe pulses through a motorized \SI{60}{\cm} long delay line (Newport M-IMS600LM). By folding the beam into six passes using hollow retro-reflectors (Newport), we obtain a maximum optical delay of \SI{12}{\ns}, i.e., we can explore the whole dynamics between subsequent pump events. The train of pump pulses is passed through an acousto-optical modulator (AOM, Gooch \& Housego 3080-125) imprinting a square-wave modulation on the pump at a frequency of \SI{1}{\MHz} and \SI{50}{\percent} duty cycle. The two beams are recombined in a second PBS and focused tightly onto the sample to an estimated focal diameter of \SI{1}{\um}. The reflected light is passed through a set of $\sfrac{\lambda}{2}$ and $\sfrac{\lambda}{4}$ waveplates in combination with a polarizer followed by an iris, allowing us to suppress the pump pulses, before the reflected probe impinges onto a fast photo diode (Femto OE-300-IN-01). We demodulate the photo diode signal on the fundamental AOM frequency using a lock-in amplifier (Zurich Instruments MFLI). Delay-dependent probe intensity time traces are extracted after averaging over \num{50} delay line passes. Its incoherent background is then removed by fitting and subtracting a polynomial (degree \numrange{20}{50}). Finally, spectra are obtained by calculating the Fourier transform of the average, background-free time trace's first derivative with respect to time (i.e., we calculate differences of subsequent voltage values, which are equidistant in time). Typical optical powers impinging on the sample were \SI{1}{\mW} for the pump and \SI{0.5}{\mW} for the probe beam for the WSe\textsubscript{2} bilayer--hBN nano-drum, and half of that for the WSe\textsubscript{2} nano-dome.

\section{\label{sec:SIlinear}Linear mass--spring--damper chain models}

Both samples investigated in this work (WSe\textsubscript{2} nano-dome and suspended WSe\textsubscript{2}-bilayer on bulk hBN) are modeled with a linear mass--spring--damper chain. Each layer in the van der Waals (vdW) material stack of $N$ layers corresponds to a mass in the linear chain, where the individual masses $m_i$, $i=1,\dotso,N$, for \SI{1}{\square\m} of material are deduced from tabulated single-layer mass densities. Interlayer forces are introduced via springs with force constants $K_i$, again deduced from tabulated interlayer force constant densities. Finally, acoustic energy losses are phenomenologically accounted for by introducing dampers of damping rates $\gamma_i$ along each spring. We assume the excitation of motion in the sample stack to be solely based on layer--layer forces $F_i(t)$, $i=1,\dotso,N-1$, induced by the pump pulse. More specifically, the coherent excitation is implemented as an ultrashort Gaussian force profile $F_i(t)$ in time driving the masses apart at time $t=0$, that is, a force $F_i$ acting on layer $i$ also acts on layer $i+1$ with opposite sign. Note that, in this way, there is no center of mass motion induced into the (vdW) material stack. The spatial dependence of the force profile is set proportional to the optical intensity of the incident laser pulse \cite{thomsen1986, lanzillotti-kimura2007}, which we simulated using a transfer matrix method and literature values for the dielectric functions of all involved materials. The full system of equations of motion then is given by
\begin{widetext}
	\begin{equation}
		\begin{split}
			m_1\ddot{x}_1&=K_1\cdot(x_2-x_1)+\gamma_1\cdot(\dot{x}_2-\dot{x}_1)+F_1,\\
			m_2\ddot{x}_2&=-K_1\cdot(x_2-x_1)+K_2\cdot(x_3-x_2)-\gamma_1\cdot(\dot{x}_2-\dot{x}_1)+\gamma_2\cdot(\dot{x}_3-\dot{x}_2)+F_2-F_1,\\
			&\vdots\\
			m_N\ddot{x}_N&=-K_{N-1}\cdot(x_N-x_{N-1})-\gamma_N\cdot(\dot{x}_N-\dot{x}_{N-1})-F_{N-1}.\label{eq:SIsystem}
		\end{split}
	\end{equation}
\end{widetext}
We rewrite the second order system of $N$ coupled equations as a first order system of $2N$ coupled equations for $x_1,\dotso,x_N,v_1,\dotso,v_N$, where $v_i\coloneqq\dot{x}_i$. Then the motion of each layer in the material stack is obtained through numeric integration using a standard numeric solver. This is performed in \textsc{Matlab} using the \texttt{ode45} routine, which implements an adaptive Runge--Kutta (4,5) method with default tolerances. Although \texttt{ode45} employs variable internal step sizes, the computed solutions were evaluated at the experimental sampling intervals of \SI{0.072}{\ps} for the nano-drum and \SI{0.5}{\ps} for the nano-dome. All times were expressed in units of \SI{1}{\ps}. One obtains excursions $x_i(t)$ of each layer relative to their positions in the undisturbed stack. Similarly, detection of motion is modeled to resemble the processes in the actual measurement. To that end, for each time $t$ the differential reflected probe signal is simulated with the assumption of the strain on each layer--layer pair contributing linearly to the change in reflectivity compared to the undisturbed case \cite{thomsen1986}. More precisely, each pair's contribution to the change in reflectivity is taken to be proportional to the instantaneous relative displacement $x_i(t)-x_{i+1}(t)$, $i=1,\dotso,N-1$, which is proportional to the instantaneous strain present on the layer pair. We then model the detected probe signal obtained from the whole layer stack with
\begin{equation}
	I_\textrm{probe}(t)=\sum_{i=1}^{N-1}\alpha_i\cdot(x_i(t)-x_{i+1}(t)),\label{eq:SIdetectedprobe}
\end{equation}
where $\alpha_i$ expresses the relative contribution of each layer pair $i$ to the detected signal. In this work, we generally set $\alpha_i=0$ if at least one of the layers $i$ or $i+1$ is not assigned to WSe\textsubscript{2}, that is, we assume detection of motion to occur solely based on strain between WSe\textsubscript{2} layers. For the remaining pairs of WSe\textsubscript{2} layers, we set $\alpha_i=\Re(E^2_i)$, where $E_i$ is the complex optical field amplitude at the position of the interface $i$ in the (vdW) stack \cite{thomsen1986}.

This optical field is obtained using the same transfer matrix simulation as for the pump. Note that the sum~\eqref{eq:SIdetectedprobe}, with the simplifications given above, takes the form of an overlap integral for the simulation of detected time-dependent changes in reflectivity through optical and acoustic transfer matrix method simulations for continuous media (applied, for example, in \cite{lanzillotti-kimura2007, lanzillotti-kimura2011}).

The final steps of the simulation are identical to the analysis applied to the experimentally measured pump--probe traces. The only additional step consists in first manually shifting the simulated signal in time to overlap temporally with the corresponding measurement as much as possible. We also apply a global amplitude scaling factor to the simulated signal, since we operate with normalized amplitudes of the optical fields and light--matter interaction strength. Note that no background is removed from the simulated data unlike for the measured signal since the simulation only reproduces components of the signal related to the coherent vibrations in the sample stack.

For the nano-drum sample consisting of WSe\textsubscript{2}-bilayer and hBN, and discussed in fig.~2 and fig.~3 of the main text, two eigenmode displacement patterns for the first and ninth longitudinal acoustic modes are shown in the inset of fig.~2(d) in the main text. They correspond to the displacement eigenvectors obtained from solving the system of equations of motion~\eqref{eq:SIsystem} with a harmonic ansatz $x_i(t)=a_i\cdot \mathrm{e}^{\textrm{i}\omega t}$, where $\omega, a_i\in\mathbb{R}$ without driving forces ($F_i\equiv0$, $i=1,\dotso,N-1$). For now, we consider the simplified, undamped case ($\gamma_i=0$, $i=1,\dotso,N-1$). In this case, one obtains the solutions of system~\eqref{eq:SIsystem} as the eigenvectors $\vb{a}=(a_i)_i$ of the $N\times N$ matrix
\begin{equation}
	\vb{A}=
	\begin{bmatrix}
		-\frac{K_1}{m_1} & \frac{K_1}{m_1} & & & \\
		\frac{K_1}{m_2} & -\frac{K_1 + K_2}{m_2} & \frac{K_2}{m_2} & & \\
		& \frac{K_2}{m_3} & -\frac{K_2 + K_3}{m_3} & \ddots & \\
		& & \ddots & \ddots & \frac{K_{N-1}}{m_{N-1}} \\
		& & & \frac{K_{N-1}}{m_N} & -\frac{K_{N-1}}{m_N}
	\end{bmatrix}.
\end{equation}
Since $\vb{A}$ is real and positive semidefinite, the eigenvalues $\omega_k^2$ are nonnegative. The mode with an eigenvalue of 0 (uniform translation) has a multiplicity of 1, so there are $N-1$ positive eigenvalues $\omega_k^2$ with corresponding eigenvectors $\vb{a}_k$ representing acoustic mode displacement patterns. If damping is considered, i.e., for at least one layer--layer pair $i$ it holds $\gamma_i>0$, damped eigenmodes with $\Im(\omega_k)>0$ can be found. For this, we introduce
\begingroup
\allowdisplaybreaks
\begin{align}
	\vb{M} &= \operatorname{diag}(m_1, m_2, \dots, m_N), \\
	\vb{C} &= 
	\begin{bmatrix}
		-\gamma_1 & \gamma_1 & & & \\
		\gamma_1 & -(\gamma_1 + \gamma_2) & \gamma_2 & & \\
		& \gamma_2 & -(\gamma_2 + \gamma_3) & \ddots & \\
		& & \ddots & \ddots & \gamma_{N-1} \\
		& & & \gamma_{N-1} & -\gamma_{N-1}
	\end{bmatrix}, \\
	\vb{K} &= 
	\begin{bmatrix}
		-K_1 & K_1 & & \\
		K_1 & -(K_1 + K_2) & \ddots & \\
		& \ddots & \ddots & K_{N-1} \\
		& & K_{N-1} & -K_{N-1}
	\end{bmatrix},
\end{align}
\endgroup
that is, the mass, damping, and stiffness matrices, respectively. Solutions to the system~\eqref{eq:SIsystem}, again without driving forces, are then found by solving the quadratic polynomial eigenvalue problem
\begin{equation}
	\left(\vb{M}\omega^2 + \vb{C}\mathrm{i}\omega + \vb{K}\right)\vb{a} = 0.\label{eq:matrixpolynomial}
\end{equation}
This is performed in \textsc{Matlab} using the \texttt{polyeig} routine after introducing units for which $\norm{\vb{M}},\norm{\vb{C}},\norm{\vb{K}} = \order{1}$.

For the WSe\textsubscript{2} nano-dome, values $m_\textrm{w}=\SI{6.0e-6}{\kg}$ and $K_\textrm{w}=\SI{8.6e19}{\N\per\m\cubed}$ are taken from \cite{vialla2020}. Based on the flake's thickness of \SI{43.5}{\nm} measured by AFM, the number of masses in the linear chain was first fixed to \num{67}, obtained from a single layer thickness of \SI{0.65}{\nm} \cite{vialla2020}, but was then slightly adjusted to \num{64} to better match the measured eigenmode frequencies. The slight missmatch in thickness may be caused by the AFM calibration. Using the damping rate $\gamma_\textrm{w}=\SI{1.26e6}{\per\s}$ obtained below for the WSe\textsubscript{2} bilayer--hBN nano-drum, the remaining damping rate $\gamma_\textrm{end}$ is manually adapted for best reproduction of the measured signal, i.e., spectrum and time trace, leading to $\gamma_\textrm{end}=\SI{1e5}{\per\s}$.

For the WSe\textsubscript{2} bilayer--hBN nano-drum, the hetero interface force constant was set to $K_\textrm{hw}=\SI{4.0e19}{\N\per\m\cubed}$ \cite{yoon2024}. Values $m_\textrm{h}=\SI{0.73e-6}{\kg}$ and $K_\textrm{h}=\SI{7.7e19}{\N\per\m\cubed}$ are taken from \cite{vialla2020}. Based on the flake's measured thickness of \SI{15.5}{\nm}, the number of hBN-assigned masses in the linear chain was first fixed to \num{47}, obtained from a single layer thickness of \SI{0.33}{\nm} \cite{vialla2020} and then slightly adjusted to \num{43} to better match the measured eigenmode frequencies. Finally, by manually adapting the remaining parameters for best reproduction of the measured signal, we found the damping rates $\gamma_\textrm{w}=\SI{1.26e6}{\per\s}$ and $\gamma_\textrm{h}=\SI{2.94e5}{\per\s}$ and we set the hetero interface damping rate to $\gamma_\textrm{hw}=\sfrac{\gamma_\textrm{h}+\gamma_\textrm{h}}{2}$.

\newpage
\FloatBarrier
\begin{widetext}
	\section{\label{sec:SIQfactor}\texorpdfstring{$Q$}{Q}-factors and free spectral range of nanoacoustic drum modes}
	
	\begin{figure*}[ht]
		\includegraphics{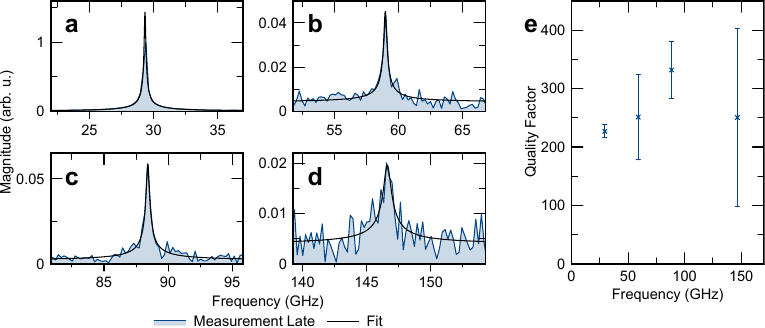}%
		\caption{\label{fig:figS2}%
			$Q$-factors of the modes in the WSe\textsubscript{2} nano-dome. They are calculated from the amplitude spectra peak positions and widths, which are extracted by fitting square roots of Lorentzian line shapes to the resonances. The measured peaks and fits are shown in panels~(a--d), where the spectra are identical to the one shown in fig.~1(g, case ``late'') of the main text. (e)~$Q$-factors obtained from the measured spectrum.}%
	\end{figure*}
	
	\begin{figure*}[ht]
		\includegraphics{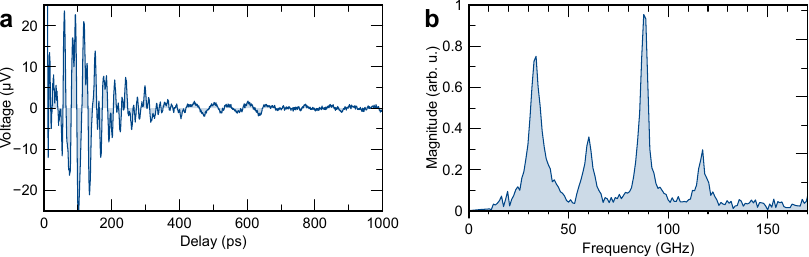}%
		\caption{\label{fig:figS3}%
			(a)~Measured differential reflectivity time trace on a planar region besides the dome. The incoherent background was removed by fitting and subtracting a polynomial of degree \num{50} in the range \qtyrange{10}{1200}{\ps}. (b)~Amplitude spectrum corresponding to (a) evaluated for \qtyrange{20}{1000}{\ps}. The $Q$-factor of the fundamental mode is found to be \num{10.4\pm1.3}, which is obtained by fitting the square root of a Lorentzian in the range \qtyrange{0}{55}{\GHz}.}%
	\end{figure*}
	
	\begin{figure*}[ht]
		\includegraphics{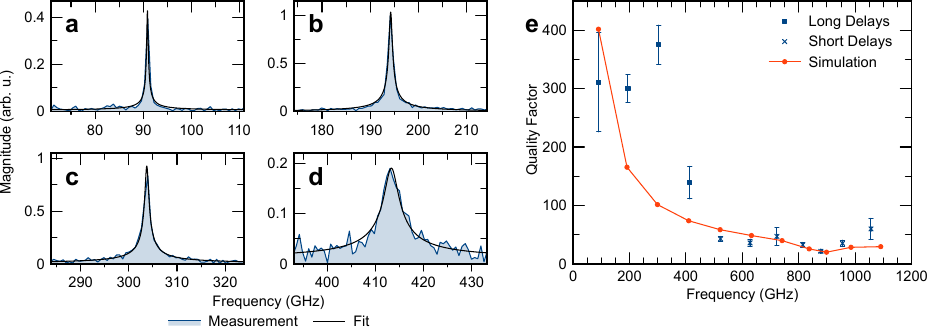}%
		\caption{\label{fig:figS4}%
			$Q$-factors of the modes in the WSe\textsubscript{2} bilayer--hBN nano-drum. They are calculated from the amplitude spectra peak positions and widths, which are extracted by fitting square roots of Lorentzian line shapes to the resonances. For the four lowest-frequency modes, an additional pump--probe measurement is performed with a longer delay range (``long delays'') to improve frequency resolution. The measured peaks and fits are shown in panels~(a--d). For the higher frequency modes, a train of peaks is fitted simultaneously to the data discussed in fig.~2(d) of the main text within the spectral range \qtyrange{450}{1100}{\GHz} (not shown). (e)~$Q$-factors obtained from the measured and simulated spectra. The $Q$-factors given for the simulated data are obtained from solving \cref{eq:matrixpolynomial}, leading to $Q_k=\sfrac{2\Im(\omega_k)}{\Re(\omega_k)}$ for the eigenmode $k$.}%
	\end{figure*}
	
	\begin{figure*}[ht]
		\includegraphics{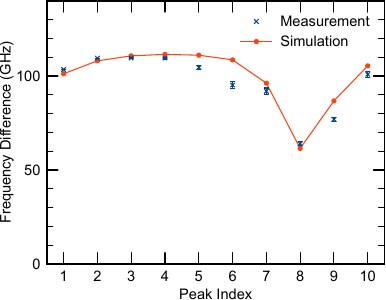}%
		\caption{\label{fig:figS5}%
			Frequency difference of subsequent resonances visible in the experimental pump--probe spectrum of fig.~2(d) in the main text (blue) and corresponding linear chain model simulation (orange). Peak positions were extracted by fitting square roots of Lorentzian line shapes to the individual resonances, as discussed in \cref{fig:figS4}. The free spectral range (FSR) at low frequencies is determined by the speed of sound in hBN and WSe\textsubscript{2} as well as the sample thickness. For the first two resonances a slight reduction in FSR results from the heavy WSe\textsubscript{2} layers terminating a chain of light hBN \cite{yoon2024}. At higher frequencies around the cut-off frequency for ZA phonons in WSe\textsubscript{2} (mode 8) the FSR reduces by almost \SI{40}{\percent} as a result of the smaller phase velocity in WSe\textsubscript{2}.}%
	\end{figure*}
	\FloatBarrier
\end{widetext}

\bibliography{local}